\documentclass[11pt]{article}

\usepackage[preprint]{acl}
\usepackage{times}
\usepackage{latexsym}

\usepackage[T1]{fontenc}
\usepackage[utf8]{inputenc}

\usepackage{microtype}

\usepackage{inconsolata}

\usepackage{graphicx}

\usepackage{amsmath}
\usepackage{amssymb}
\usepackage{mathtools}
\usepackage{amsthm}
\usepackage{amsfonts}
\usepackage{bm}

\usepackage{booktabs}
\usepackage{multirow}

\usepackage{float}
\usepackage{subcaption}

\usepackage{pifont}
\usepackage{xcolor}
\usepackage{siunitx}
\usepackage{wrapfig}

\theoremstyle{plain}

\theoremstyle{definition}

\theoremstyle{remark}

\title{Revisiting RAG Retrievers: An Information Theoretic Benchmark}

\author{Wenqing Zheng \and Dmitri Kalaev \and Noah Fatsi \and Daniel Barcklow \\
  \and Owen Reinert \and Igor Melnyk \and Senthil Kumar \and C. Bayan Bruss \\
  Capital One \\
  \texttt{wenqing.zheng@capitalone.com}}

\begin{document}
\maketitle
% Although the framework is task-agnostic and applicable to any setting where a ground-truth answer can construct a pseudo chunk distribution, our empirical study focuses on multi-hop question answering --- a demanding setting where retriever complementarity is especially pronounced. 
\vspace{-0.4em}
\begin{abstract}
\vspace{-0.2em}
Retrieval-Augmented Generation (RAG) systems rely critically on the retriever module to surface relevant context for large language models. Although numerous retrievers have recently been proposed, each built on different ranking principles such as lexical matching, dense embeddings, or graph citations, there remains a lack of systematic understanding of how these mechanisms differ and overlap. Existing benchmarks primarily compare entire RAG pipelines or rely on aggregate ranking metrics (Recall, MRR, nDCG) that assume item-level independence, providing little guidance on redundancy or complementarity between retrievers. This work presents MIGRASCOPE, a Mutual Information based RAG Retriever Analysis Scope. We introduce principled metrics grounded in information theory to quantify retrieval quality, redundancy/synergy, and marginal contribution. Although the framework is designed to be applicable to any setting where a ground-truth answer can construct a pseudo chunk distribution, the present empirical study focuses on multi-hop question answering \textemdash a demanding setting where retriever complementarity is especially pronounced. On four corpora and thirteen SOTA retrievers spanning lexical, dense, and graph-based families, we show that (i) the MI-based divergence metric captures retriever quality dimensions not reflected by Recall or MRR, (ii) graph-based retrievers are redundant with each other while BM25 remains informationally complementary, and (iii) in the multi-hop QA setting, a carefully chosen ensemble guided by MI-based attribution can outperform any single retriever. We position these findings as task-specific insights and provide a reusable framework for retriever analysis that practitioners can re-estimate on their own target domains. 

\end{abstract}
% ; we include a preliminary single-hop transfer experiment to probe generalizability (\S\ref{sec:transfer}).
% Retrieval-Augmented Generation (RAG) systems rely critically on the retriever module to surface relevant context for large language models. Although numerous retrievers have recently been proposed, each built on different ranking principles such as lexical matching, dense embeddings, or graph citations, there remains a lack of systematic understanding of how these mechanisms differ and overlap. Existing benchmarks primarily compare entire RAG pipelines or introduce new datasets, providing little guidance on analyzing retriever interactions themselves. Those that do compare retrievers directly use a limited set of evaluation tools which fail to capture complementary and overlapping strengths. This work presents MIGRASCOPE, a Mutual Information based RAG Retriever Analysis Scope. We revisit state-of-the-art retrievers and introduce principled metrics grounded in information and statistical estimation theory to quantify retrieval quality, redundancy, synergy, and marginal contribution. We further show that, in the evaluated multi-hop QA setting, a carefully chosen ensemble of retrievers can outperform any single retriever. We leverage the developed tools over major RAG corpora to provide unique insights on contribution levels of the state-of-the-art retrievers. Our findings provide a fresh perspective on the structure of modern retrieval techniques and a task-specific framework for analyzing retriever interactions in robust and efficient RAG systems.

\section{Introduction}

Retrieval-Augmented Generation (RAG) has emerged as a dominant paradigm for knowledge-intensive NLP tasks \cite{lewis2020retrieval}. By retrieving relevant external context to inform large language models (LLMs), RAG systems dramatically improve factual accuracy and reduce hallucination \cite{shuster2021retrieval}. The retriever module is the core component responsible for selecting which evidence is passed to the generator, and its effectiveness directly limits downstream performance. A rich ecosystem of retrievers has flourished, including lexical-based methods such as BM25 \cite{robertson2009probabilistic}, dense embedding models trained for semantic similarity \citep{karpukhin2020dense}, graph-based retrievers designed for multi-hop and global reasoning \citep{edge2024local}, and increasingly hybrid retrieval systems \cite{sarmah2024hybridrag}. Each introduces distinct assumptions and ranking mechanisms, yet the field lacks a unified framework to compare them rigorously.
Many existing benchmarks focus on new datasets or evaluation metrics applied to full RAG pipelines. While these studies provide valuable insights into overall system quality, they rarely isolate retrieval performance.
Some retriever-specific evaluation work has been done, covering heterogeneous domain generalization \cite{thakur2021beir}, massive embedding leaderboards \cite{muennighoff2023mteb}, and complex-objective tasks \cite{wang2024birco}. However these frameworks focus on aggregate ranking-based metrics such as recall, MRR, and nDCG which assume independence between retrieved items and thus fail to explain how different retrievers complement or supersede each other.
Recent ensemble-RAG efforts mostly concentrate on hierarchical integration of the full pipeline without disentangling the unique contributions of the retriever itself. As a result, practitioners have limited principled tools for quantifying which retrievers are strong, which are redundant, and which offer complementary value on a given dataset. We argue that revisiting the retriever module with dedicated evaluation tools is essential to unlocking the next phase of RAG progress.

To address these gaps, we introduce MIGRASCOPE, an information-theoretic framework for analyzing and benchmarking retrievers in RAG. We model each retriever's ranking signal as a noisy view of an underlying chunk probability distribution conditioned on ground-truth answers. Building on Mutual Information (MI), we define a retriever quality score that computes the diverging level from the presumed ideal chunk distribution. This score generalizes across lexical, dense, and graph-based retrieval mechanisms, enabling principled comparison and calibration.

We further develop MI-grounded synergy and redundancy metrics to quantify how multiple retrievers interact - whether they contribute complementary evidence or duplicate signals - and provide visual diagnostics to reveal overlap structure. These tools support task-specific marginal contribution estimation and guide retriever analysis and shortlist construction within a target retrieval setting.

Finally, we translate these analyses into practice with a lightweight ensemble that uses MI-based attribution to select and weight retrievers with minimal supervision. Applied across major RAG corpora and a broad suite of state-of-the-art (SOTA) retrievers, the framework yields new insights into contribution patterns, redundancy, and robustness, and the ensemble consistently outperforms strong single retrievers in our evaluated multi-hop QA setting. This paper makes the following contributions:
\begin{itemize}
\vspace{-0.6em}
\item Introduce MIGRASCOPE, an MI based retriever quality score that normalizes across ranking paradigms (lexical, dense, graph) to quantify retrieval relevance and utility.
\vspace{-0.6em}
\item Define and visualize MI-grounded synergy and redundancy metrics for multiple retrievers, revealing complementary and overlapping signal structure within ensembles.
\vspace{-0.6em}
\item Develop a retriever ensemble framework with MI-based attribution to estimate each retriever's marginal contribution within a task-specific retrieval setting.
\vspace{-0.6em}
\item Benchmark a broad set of graph-RAG and standard RAG retrievers using the proposed methodology, yielding new insights on contribution patterns, robustness, and design guidance for multi-hop QA RAG systems.
\end{itemize}

\vspace{-0.8em}
Our results suggest a need to shift community focus from endlessly inventing new retrievers toward understanding how existing ones differ, overlap, and cooperate within a target retrieval environment. Through an information-theoretic lens, we uncover structure and simplicity hidden beneath the diversity of modern retrieval methods, paving the way for more reliable and interpretable RAG systems.

\begin{figure*}[h]
    \centering
    \renewcommand{\thefigure}{A}
    \includegraphics[trim=1.5em 0.8em 1.7em 0.2em, clip,width=0.97\linewidth]{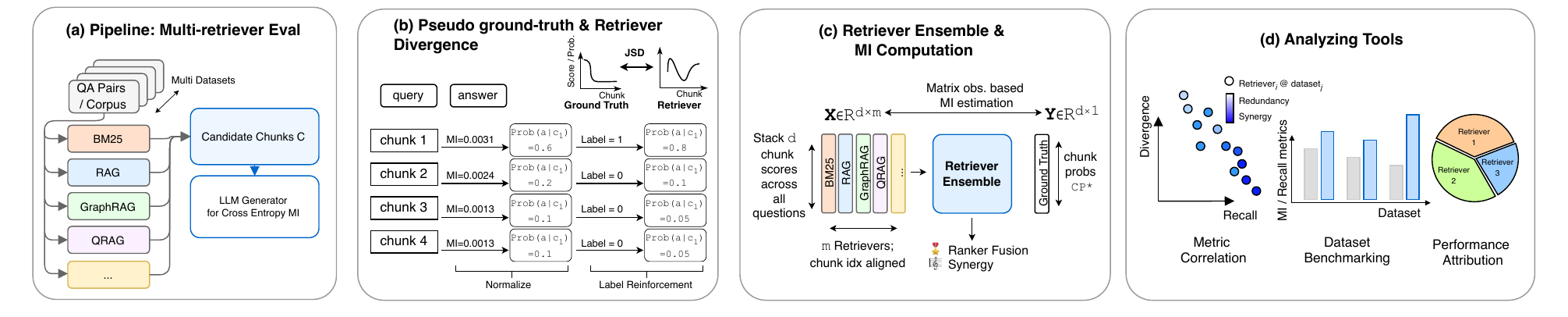}
    \caption{The proposed MIGRASCOPE analyzing framework}
    \label{fig:arch}
\end{figure*}

\vspace{-0.4em}
\section{Related Work}
\label{sec:related_work}
\vspace{-0.6em}

Our work is positioned at the intersection of three active research areas: RAG benchmarking, RAG ensemble methods, and the application of information-theoretic principles to large language models.

\vspace{-0.8em}
\subsection{RAG Benchmarking}
Recent work on RAG evaluation has primarily focused on establishing new datasets, defining system-level metrics, or analyzing operational costs, rather than systematically comparing the retrievers themselves. Several benchmarks have been introduced to test RAG systems on complex tasks, such as multi-hop reasoning over structured documents \citep{khang2025crest} or domain-specific GraphRAG applications \citep{xiao2025graphrag}. Others have focused on creating datasets for specific industrial domains, such as finance \citep{wang2024omnieval}, or on proposing new evaluation methodologies, like assessing LLM robustness and information integration \citep{chen2024benchmarking} or developing explainable metrics for industrial settings \citep{friel2024ragbench}.

To evaluate retrievers in isolation, the Information Retrieval community has developed rigorous benchmarks covering heterogeneous domain generalization \citep{thakur2021beir}, massive embedding leaderboards \citep{muennighoff2023mteb}, and complex user objectives \citep{wang2024birco}. However, these frameworks typically rely on aggregate ranking metrics such as nDCG and Recall. While effective for static leaderboards, these metrics treat document utility as additive, implicitly assuming independence between retrieved items. Consequently, they fail to quantify redundancy or measure the conditional information gain required to understand how distinct retrievers complement or supersede one another in an ensemble.

While valuable, these contributions often treat the retriever as a component within a larger pipeline. The focus remains on dataset design or system-level assessment. Other studies analyze important, but orthogonal, aspects. For instance, \citet{lin2024operational} provide practical guidance on the operational trade-offs of different retrieval infrastructures (e.g., HNSW vs.\ inverted indexing) but do not offer prescriptive insights on retriever algorithm selection. Similarly, \citet{zhang2025comparative} compare different LLMs as dense retrievers, but limit their analysis to this single family.

Across these studies, there remains a significant gap in understanding the fundamental differences, redundancies, and complementary strengths of the various retriever families (e.g., dense, sparse, hybrid, and graph-based). Our work directly addresses this gap by proposing a framework to benchmark the retrievers themselves.

\vspace{-0.4em}
\subsection{RAG Ensemble Methods}
\vspace{-0.4em}
As RAG systems have matured, ensemble and multi-retriever architectures have become increasingly common. Many of these efforts focus on sophisticated pipeline orchestration. For example, \citet{zhang2025levelrag} propose a hierarchical system with a planner that orchestrates multiple low-level searchers, while \citet{wu2025composerag} introduce a modular framework that composes distinct steps like question decomposition and retrieval. Other approaches integrate heterogeneous data sources, such as combining web search with knowledge graphs \citep{xie2024weknow} or creating adaptive hybrid systems for specific domains like law \citep{kalra2024hypa}.

A separate line of work seeks to optimize the information passed to the generator. This includes methods for decoupling retrieval and generation representations \citep{yang2025heterag} or explicitly balancing relevance and semantic diversity in the retrieved document set \citep{rezaei2025vendi}. \citet{chen2025revisiting} explicitly analyze ensembles through an information-theoretic lens, arguing that they increase the total useful information. However, their work stops short of providing an operational recipe for selecting or weighting retrievers. These studies primarily offer architectural solutions or in-domain systems, but do not provide a principled, data-driven method for understanding \emph{which} retrievers to ensemble or \emph{how} to combine them based on their quantified marginal contribution.

\vspace{-0.4em}
\subsection{Information-Theoretic Analysis in RAG}
\vspace{-0.4em}
Information theory provides a powerful, formal toolkit for quantifying concepts like uncertainty, relevance, and redundancy, which are central to RAG. Recently, these tools have been applied to various parts of the RAG pipeline. For instance, \citet{zhu2024information} use an information bottleneck formulation to filter noise from retrieved contexts. \citet{liu2024pointwise} propose using Pointwise Mutual Information (PMI) as a probabilistic gauge for RAG performance, while others have used divergence metrics to measure attribution \citep{li2025attributing} or detect hallucinations \citep{halperin2025prompt}. These methods successfully apply information-theoretic concepts to \emph{diagnose} or \emph{optimize} a specific RAG component, such as context pruning \citep{deng2025influence} or detecting data memorization \citep{anh2025repcs}.

However, this body of work has not yet been directed at the problem of \emph{retriever evaluation}. While \citet{yumucsak2025information} outline a theoretical framework for RAG information flow, it lacks the empirical benchmarking to reveal how different SOTA retrievers actually perform under these metrics. To our knowledge, no existing work uses mutual information and information divergence to create a comparative framework that quantifies the quality, redundancy, and marginal contribution of distinct retriever families.

In summary, the field currently lacks a unified framework for comparing retriever modules directly, a principled method for designing retriever ensembles based on redundancy, and an empirical application of information theory to benchmark retriever contributions. Our work, MIGRASCOPE, is designed to fill these precise gaps.

\vspace{-0.4em}
\section{Analytical Framework}
\vspace{-0.4em}

\subsection{Preliminaries: GraphRAG Formulations}
Let a query be $q\in\mathcal{Q}$, a candidate chunk set be $C=\{c_1,\dots,c_N\}$, and an answer be $a\in\mathcal{A}$. A retriever $r$ maps $q$ to a scored or ranked subset $\mathcal{C}_q\subseteq C$ as the context.
The indices of chunks subset $\mathcal{C}_q$ are denoted as $\mathcal{I}_q$.
The retriever gives each chunk a score, combining into a scored list $\{(c_j,s_j)\}_{j\in\mathcal{I}_q}$.
A generator then produces the answer $a$ conditioned on $(q,\mathcal{C}_q)$.
We formulate a simplified version of RAG and GraphRAGs here. More details are described in Appendix~\ref{appendix:grag formulations}.

\noindent\textbf{Vanilla RAG.} The retrieval $\mathcal{C}_q$ relies solely on vector similarity to select the top-$k$ chunks:
\begin{small}
\begin{equation}
\mathcal{C}_q = \text{Top}_k(\{c_j\}_{j=1}^N, \mathbf{Sim}(\mathbf{Embed}(q), \mathbf{Embed}(c_j)))
\end{equation}
\end{small}

\noindent\textbf{GraphRAG (General Form).} GraphRAG introduces a knowledge graph $G$ constructed from the corpus. Retrieval is a two-stage process: graph anchoring and context mapping.
\begin{gather}
G = \mathbf{Construct}(C) \\ \mathcal{C}_q = \mathbf{Map}_{G \to C}(\mathbf{Anchor}(G, q))
\end{gather}

\subsection{Information-Theoretic View of RAG}

\paragraph{Mutual information (MI).}
For random variables $X,Y$ with joint $p(x,y)$ and marginals $p(x),p(y)$,

\begin{multline}
I(X;Y)=\mathbb{E}_{p(x,y)}\!\left[\log\frac{p(x,y)}{p(x)\,p(y)}\right]\\
=H(X)-H(X\mid Y)=H(Y)-H(Y\mid X).
\label{eq:mi}
\end{multline}
MI measures the reduction in uncertainty of one variable given the other; it is zero iff \(X\) and \(Y\) are independent.

\paragraph{Divergence.}
In this work we apply Jensen--Shannon divergence (\(\mathrm{JSD}\)) which is symmetric, bounded in \([0,\log 2]\), and well-defined even with disjoint supports.
\begin{multline}
\mathrm{KL}(P\parallel Q)=\mathbb{E}_{p}\!\left[\log\frac{p}{q}\right], \\ \mathrm{JSD}(P,Q)=\tfrac12\,\mathrm{KL}\!\big(P\Vert M\big)+\tfrac12\,\mathrm{KL}\!\big(Q\Vert M\big),\\
M=\tfrac12(P+Q).
\label{eq:jsd}
\end{multline}

\paragraph{Pointwise mutual information in RAG.}
Given query $q$ and a random multiset of retrieved chunks $\mathcal{C}_q$, define the pointwise mutual information w.r.t.\ the answer $a$ as
\vspace{-0.4em}
\begin{equation}
\mathrm{PMI}(a;\mathcal{C}_q\mid q)=\log\frac{p(a\mid \mathcal{C}_q,q)}{p(a\mid q)}.
\end{equation}
\vspace{-0.4em}
Exact evaluation is intractable because the answer $a$ is generative with a combinatorially large support. In practice we approximate with a fixed ground-truth answer $a$ from the dataset:
\begin{equation}
\widehat{\mathrm{PMI}}(a;\mathcal{C}_q\mid q)\approx \log p(a\mid \mathcal{C}_q,q)-\log p(a\mid q).
\end{equation}
We estimate $\log p(a\mid \cdot)$ with the LLM token-level cross-entropy (log perplexity). For a tokenization $a=(t_1,\dots,t_T)$ and model condition $Z$,
\begin{small}
\begin{multline}
\log p(a\mid Z)=\sum_{t=1}^{T}\log p_\theta(t_t\mid t_{<t},Z) \\
\Longleftrightarrow\quad
\mathrm{CE}(a\mid Z)=-\frac{1}{T}\sum_{t=1}^{T}\log p_\theta(t_t\mid t_{<t},Z).
\end{multline}
\end{small}
\vspace{-0.4em}

where $\theta$ represents the generative LLM parameters. Large PMI between the answer $a$ and context $\mathcal{C}_q$ corresponds to large perplexity drop under retrieved context $Z=(q,\mathcal{C}_q)$ relative to no context $Z=(q)$. PMI, then, quantifies how much a retrieved set of chunks $C_q$, for a given question $q$, reduces uncertainty in the ground-truth answer $a$.

\subsection{Pseudo Ground-Truth Chunk Probability}

We define a pseudo ground truth probability over the retrieved chunks that reflects how well each chunk supports the known answer. Related chunk attribution techniques utilize LLM cross-entropy to gauge chunk importance, either by calculating Information Bottleneck scores \cite{zhu2024information} or by measuring the performance degradation when a specific chunk is removed from the context \cite{deng2025influence}. However, such methods often introduce undesirable complexity. Ablation-based approaches suffer from computational overhead and bias induced by chunk order, as LLMs have been shown to disproportionately favor the first and last chunks in a sequence \cite{liu2024pointwise}. Meanwhile, Information Bottleneck methods require tuning sensitive hyperparameters and optimizing auxiliary reconstruction objectives. To avoid these pitfalls, we apply a simple, independent scoring approach to measure the chunk self-attribution score. Thus we have the pseudo ground-truth chunk probability $\mathrm{CP}^\ast$:

\vspace{-0.4em}
\begin{equation}
\label{eq:CPast}
\mathrm{CP}^\ast(c\mid q,a)\;=\mathrm{Softmax}_{\mathcal{C}_q}(\log p_\theta(a\mid q,c))
\end{equation}
\vspace{-0.4em}
where $\mathrm{Softmax}_{\mathcal{C}_q}$ normalizes the chunk probabilities based on the logits over the subset of retrieved chunks $\mathcal{C}_q$.

\paragraph{Golden chunk reinforcement.}
If the dataset provides golden supporting chunks $G_q\subseteq \mathcal{I}_q$, we optionally reinforce them by a scalar $\gamma>1$, i.e.\ amplify the dataset labeled ground truth chunks then re-normalize
\footnote{When golden chunk support exists, the chunk distribution is ideally one-hot. The $\mathrm{CP}^\ast$ softens the hard distribution. For better alignment with dataset labels, we empirically pick large $\gamma$.}:
\begin{tiny}
\begin{equation}
\label{eq:CPasttilde}
\widetilde{\mathrm{CP}}^\ast(c\mid q,a)=
\frac{\mathbf{1}[c\in G_q]\gamma\cdot \mathrm{CP}^\ast(c\mid q,a)
+\mathbf{1}[c\notin G_q]\cdot \mathrm{CP}^\ast(c\mid q,a)}
{\sum_{j\in\mathcal{I}_q}\big(\mathbf{1}[c_j\in G_q]\gamma+\mathbf{1}[c_j\notin G_q]\big)\mathrm{CP}^\ast(c_j\mid q,a)}.
\end{equation}
\end{tiny}
We use $\widetilde{\mathrm{CP}}^\ast$ as the default pseudo target distribution to produce the following retriever analysis.

\subsection{Retriever Divergence Score}
Once we have the pseudo ground-truth chunk probability \texorpdfstring{$\mathrm{CP}^\ast$}{CP*} as the standard to compare with, we can easily measure the quality of any retriever $r$ by comparing the scores produced by this retriever and \texorpdfstring{$\mathrm{CP}^\ast$}{CP*}. We first convert the retriever scores $s_r(c)$ to probabilities:
\begin{equation}
p_r(c\mid q)=\frac{\exp(s_r(c)/\tau)}{\sum_{j\in\mathcal{I}_q}\exp(s_r(c_j)/\tau)}
\end{equation}
where $\tau>0$ is a temperature-softmax normalization. The quality of the retriever $r$ can be defined by the divergence of $r$ from the pseudo target:
\begin{small}
\begin{equation}
\label{eq:div}
\mathrm{Div}(r)\;=\;\frac{1}{|\mathcal{Q}|}\sum_{q\in\mathcal{Q}}
\mathrm{JSD}\!\left(p_r(\cdot\mid q)\;\Big\|\;\widetilde{\mathrm{CP}}^\ast(\cdot\mid q,a_q)\right).
\end{equation}
\end{small}

\vspace{-0.6em}
\subsection{Ensembling Multiple Retrievers}
Assume $m$ retrievers $\{r_i\}_{i=1}^m$ produce scores $\{s_i(c)\}_{i=1}^m$. There are multiple ways that these retrievers can be fused. We list and compare 12 ways to merge the rankers and show results in Section~\ref{sec:results}. We do not aim to claim novelty on the ranker fusion approaches; we leverage existing approaches as detailed in Appendix~\ref{appendix:ranker fusion}.

\subsection{Marginal Contributions and Redundancy Analysis}
In the attempt to benchmark and analyze different retrievers, we utilize well-established tools from statistical signal processing to quantify the contribution and redundancy levels.

\paragraph{Setup.}
Given a batch of queries $\{q_i\}_{i=1}^{|\mathcal{Q}|}$, we stack the retriever's scores and \texorpdfstring{$\mathrm{CP}^\ast$}{CP*} to obtain observation $X$ and estimation target $Y$.

For each query, we use $\widetilde{\mathrm{CP}}^\ast$ to produce a list of top $K$ chunk scores, and concatenate them into target vector $Y\in\mathbb{R}^{K|\mathcal{Q}|}$

On the other hand, each of the $m$ retrievers also produces scores for these top $K$ chunks. We stack them into the observation matrix $X\in\mathbb{R}^{K|\mathcal{Q}|\times m}$.

\paragraph{Utility.}
Define the set function
\begin{equation}
\label{eq:util}
F(S)=I\big(Y;X_S\big),
\end{equation}
which measures predictive information about $Y$ contained in the subset $S$ of retrievers. This is distinct from Fisher information.

\paragraph{Independent marginal contribution.}
Denote the collection of $m$ retrievers as $\mathcal{R}$. For retriever $i$, its independent marginal contribution is:
\begin{multline}
C_i\;\stackrel{\text{def}}{=} \;I\big(Y;X_i\mid X_{\mathcal{R}/\{i\}}\big)\\
\;=\;H\big(Y\mid X_{\mathcal{R}/\{i\}}\big)-H\big(Y\mid X_{\mathcal{R}}\big),
\end{multline}

where $X_{\mathcal{R}/\{i\}}$ denotes all columns of $X$ excluding the column corresponding to retriever $i$, $X_{\mathcal{R}}$ denotes the full observation matrix. Large $C_i$ then captures the unique information conveyed by retriever $i$. Values near zero indicate redundancy given the others.

\paragraph{Shapley value for fair attribution.}
Define Shapley values over the coalition game $(\{1,\dots,m\},F)$:
\begin{tiny}
\begin{equation}
\phi_i=\sum_{S\subseteq \{1,\dots,m\}\setminus\{i\}}
\frac{|S|!\,(m-|S|-1)!}{m!}\left[F(S\cup\{i\})-F(S)\right].
\end{equation}
\end{tiny}
Shapley values fairly allocate the total utility $F(\{1{:}m\})-F(\varnothing)$ across retrievers by $\Sigma_{i=1}^m \phi_i = F(\{1{:}m\})-F(\varnothing)$. They average marginal information gains over all insertion orders, which resolves order dependence.

\paragraph{Redundancy vs synergy.}
For any pair $(i,j)$, define the interaction information

\vspace{-0.4em}
\begin{equation}
\label{eq:redundancy}
\mathrm{II}(Y;X_i;X_j) \;=\; I(Y;X_i) - I\big(Y;X_i \mid X_j\big).
\end{equation}
\vspace{-0.4em}
Positive $\mathrm{II}$ indicates redundancy; negative $\mathrm{II}$ indicates synergy, i.e., complementary information that emerges only jointly.
Accordingly, we define a surrogate distance between retrievers $r_i$ and $r_j$ as

\vspace{-0.4em}
\begin{equation}
\label{eq:dist def}
d(r_i,r_j) = \exp\!\big(-\mathrm{II}(Y;X_i;X_j)\big).
\end{equation}
\vspace{-0.4em}

We further leverage this score to position each retriever on a \textit{redundancy--synergy spectrum}, and perform clustering analysis to reveal the structural grouping of retrievers. Results are presented in Section~\ref{sec:spectrum}.

\paragraph{Retriever Selection Principle.}
We apply the tractable redundancy/synergy scores in Eq.~\eqref{eq:redundancy} to decide whether to include certain RAG retriever for ensemble. We follow the principle that multicollinearity harms estimation stability but independence helps. For any sources $X_1, X_2$ and target $Y$, $I(Y; X_1,X_2)=I(Y;X_1)+I(Y;X_2|X_1)$. If $X_2$ is redundant given $X_1$, then $I(Y;X_2|X_1)=0$. If $X_2$ is complementary, this conditional term is positive, so the pair contains strictly more information than either alone.

\subsection{Estimating Entropy and Mutual Information from Vector Observations}
\label{sec:est MI under vec}

Assuming X and Y are jointly Gaussian yields a closed-form mutual information (MI) estimator (see Appendix~\ref{appendix:gaussian assumption}). However, this assumption is theoretically tenuous in our setting and, empirically, leads to systematic underestimation of MI by failing to capture nonlinear dependencies between the variables.

Therefore, we accept the distributions of $X$ and $Y$ as unknown, then train a predictive model $\hat{p}(y\mid x)$ and compute $\widehat{H}(Y\mid X)=\tfrac{1}{n}\sum_i H(\hat{p}(\cdot\mid x_i))$ for discrete $Y$, or approximate continuous $Y$ with a Gaussian residual model where $\widehat{H}(Y\mid X)=\tfrac{1}{2}\log(2\pi e\,\widehat{\sigma}^2_{\mathrm{res}})$. We use XGB for regression.

\begin{table*}[t]
    \centering
    \resizebox{0.8\textwidth}{!}{
        \begin{tabular}{l ccc ccc ccc ccc}
            \toprule[2pt]
            \multirow{2}{*}{\textbf{Method}} & \multicolumn{3}{c}{\textbf{2Wiki}} & \multicolumn{3}{c}{\textbf{HotpotQA}} & \multicolumn{3}{c}{\textbf{MuSiQue}} & \multicolumn{3}{c}{\textbf{TriviaQA}} \\
            \cmidrule(lr){2-4} \cmidrule(lr){5-7} \cmidrule(lr){8-10} \cmidrule(lr){11-13}
             & \textbf{Rec} & \textbf{MRR} & \textbf{Div} & \textbf{Rec} & \textbf{MRR} & \textbf{Div} & \textbf{Rec} & \textbf{MRR} & \textbf{Div} & \textbf{Rec} & \textbf{MRR} & \textbf{Div} \\
            \midrule

            \textbf{G-window} & 0.935 & 0.959 & 0.110 & 0.871 & 0.916 & 0.118 & 0.822 & 0.806 & 0.125 & 0.891 & 0.922 & 0.102 \\
            \textbf{G-threshold} & 0.936 & 0.960 & 0.120 & 0.921 & 0.924 & 0.121 & 0.725 & 0.805 & 0.122 & 0.944 & 0.922 & 0.101 \\
            \textbf{G-hierarchical} & 0.934 & 0.958 & 0.117 & 0.872 & 0.916 & 0.120 & 0.724 & 0.803 & 0.129 & 0.881 & 0.914 & 0.098 \\
            \textbf{G-naive} & 0.957 & 0.934 & 0.122 & 0.876 & 0.920 & 0.125 & 0.729 & 0.808 & 0.134 & 0.892 & 0.933 & 0.144 \\
            \textbf{G-community} & 0.935 & 0.959 & 0.125 & 0.873 & 0.917 & 0.123 & 0.727 & 0.807 & 0.130 & 0.891 & 0.919 & 0.119 \\
            \textbf{G-relationship} & 0.934 & 0.958 & 0.128 & 0.873 & 0.917 & 0.124 & 0.728 & 0.807 & 0.129 & 0.845 & 0.922 & 0.127 \\
            \textbf{G-global} & 0.937 & 0.961 & 0.125 & 0.880 & 0.922 & 0.128 & 0.726 & 0.805 & 0.131 & 0.835 & 0.955 & 0.121 \\
            \textbf{HippoRAG} & 0.931 & 0.960 & 0.142 & 0.892 & 0.921 & 0.128 & 0.833 & 0.810 & 0.120 & 0.921 & 0.920 & 0.119 \\
            \textbf{RAG} & 0.911 & 0.921 & 0.133 & 0.851 & 0.889 & 0.130 & 0.711 & 0.768 & 0.159 & 0.860 & 0.891 & 0.125 \\
            \textbf{LLM} & 0.913 & 0.944 & 0.131 & 0.824 & 0.908 & 0.138 & 0.807 & 0.801 & 0.155 & 0.882 & 0.914 & 0.129 \\
            \textbf{QRAG} & 0.889 & 0.919 & 0.147 & 0.713 & 0.792 & 0.133 & 0.681 & 0.753 & 0.166 & 0.829 & 0.881 & 0.144 \\
            \textbf{LightRAG} & 0.785 & 0.863 & 0.155 & 0.507 & 0.644 & 0.162 & 0.563 & 0.659 & 0.199 & 0.877 & 0.820 & 0.157 \\
            \textbf{BM25} & 0.486 & 0.605 & 0.252 & 0.620 & 0.739 & 0.238 & 0.415 & 0.507 & 0.302 & 0.420 & 0.569 & 0.289 \\
            \bottomrule[2pt]
        \end{tabular}
    }
    \caption{RAG Retriever Benchmarking with MI metric}
    \label{tab:rag_benchmark}
\end{table*}

\section{Results}
\label{sec:results}

\subsection{Benchmarking RAG Retrievers}

In this work, we focus on the effect of retriever mechanisms rather than encoder, hence the encoder remains fixed and same for all RAG approaches in our work. We use BGE-M3 \cite{chen2024bge} indiscriminatively.
We benchmark a dozen of SOTA RAG retrievers, including chunk similarity based (vanilla) retriever, lexical retrievers, graph retrievers, and chunk decomposition based retrievers, etc. For the seven G-series graph retrievers, the encoder and indexing backbone are shared, while the graph construction differs: windowed co-occurrence, thresholded co-occurrence, hierarchical provenance, naive co-occurrence, community, relationship-global, and global co-occurrence. We additionally compare with HippoRAG \cite{gutierrez2025rag} and LightRAG \cite{guo2024lightrag}.
% \footnote{Our codes are available at \url{https://github.com/CapitalOne-Research/Migrascope}}

\subsection{Datasets and Split}
We evaluate the retrievers on four multi-hop QA corpora: HotpotQA, MuSiQue, 2WikiMultiHopQA, TriviaQA.
For each corpus, we uniformly sample $1{,}000$ QA pairs.

Benchmarking results for these retrievers are presented in Table~\ref{tab:rag_benchmark}, which shows the traditional metrics (Recall@1, MRR) as well as the newly derived divergence metric from Eq.~\eqref{eq:div}.
Next, we leverage the proposed tools to address four key research questions.

\begin{itemize}
    \item RQ1: How does the new MI based metric align with previous MRR/Recall metrics?
    \item RQ2: How does the selection of hyper-parameters impact MI?
    \item RQ3: Is it possible to combine multiple weak retrievers into a stronger retriever, and how are they contributing individually?
    \item RQ4: Among the SOTA retrievers, how redundant are they?
\end{itemize}

More results are presented in Appendix~\ref{sec:more_res}, including variances, computational costs, ranking robustness, case study of MI/Recall metric differences, MIGRASCOPE analyzing workflow, etc.

\subsection{Correlation Analysis (\textbf{RQ1})}
\begin{figure*}[t]
    \centering
    \renewcommand{\thefigure}{1}
    \includegraphics[width=0.95\linewidth]{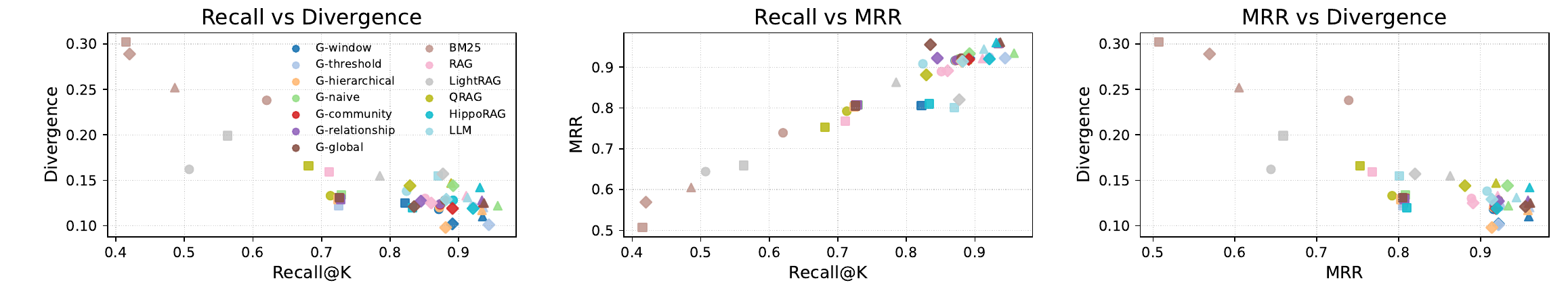}
    \caption{Metric Correlations Across Datasets, Legend by Retriever}
    \label{fig:1_div_corr}
\end{figure*}

This experiment aims to check the level of alignment between the proposed divergence metric and traditional metrics. For each (retriever, dataset) pair, we plot the divergence score against standard evaluation metrics (Recall@$1$, MRR) in Figure~\ref{fig:1_div_corr}.

The first observation is that recall and \textit{Div} scores are negatively correlated in general (strong retrievers tend to have lower divergence).
There are also systematic deviations, as multiple retrievers achieve similar recall but differ significantly in divergence.
This confirms the complementary nature of our divergence metric and its ability to diagnose failure modes not captured by ranking accuracy alone.

These deviations also affect practical selection decisions. On 2Wiki, rank-based metrics among strong graph retrievers can be contradictory, while \textit{Div} selects the retriever that later contributes more strongly in the best ensemble. On MuSiQue, several graph variants are marginally stronger than G-hierarchical under Recall/MRR, yet \textit{Div} identifies G-hierarchical as the better ensemble component. 
% Appendix~\ref{appendix:decision_examples} summarizes these concrete decision differences and the corresponding recall drops after replacement.

\subsection{Sensitivity of MI Hyperparameters (\textbf{RQ2})}
\label{sec:hparam}
We sweep top-$K$, anchor construction, and golden-chunk reinforcement strength $\gamma$; the full plot is deferred to Appendix~\ref{appendix:hparam_robustness} to conserve space. The same qualitative pattern remains: larger $\gamma$ makes \textit{Div} more recall-like, while larger top-$K$ reduces Pearson alignment because retrievers increasingly differ from the pseudo target away from the top evidence. Importantly, retriever ordering is more stable than raw Pearson values alone suggest. On HotpotQA, the Spearman correlation of \textit{Div}-induced retriever rankings across the sweep stays within $[0.748, 0.910]$, with per-setting means between $0.841$ and $0.852$ (Table~\ref{tab:spearman_hparam}). This supports using \textit{Div} primarily as a relative diagnostic for model selection rather than an absolute standalone score.

\subsection{Retriever Ensemble and Shapley Contribution Values (\textbf{RQ3})}

We conduct a series of experiments in the attempt to find a subset of combined retrievers outperforming any single strongest retriever, as well as visualizing the synergy across different retrievers.

Our retriever ensemble is based on the divergence metric. For every given dataset, we first run retrieval procedure with every retriever to get a large top-K chunk list for 1000 QA pairs. Then we select a list of retriever subset to perform ensemble search in the unified search space.
% , then the best retriever ensemble algorithm is found within that space, using the 20\% training QA pairs. Finally, the ensembled retriever is evaluated on the test 80\% QA pairs. 
The best fusion rule is selected on the \textbf{20\% training split} by maximizing held-out MI on a nested 5-fold cross-validation within that split; no test labels are touched during fusion-rule selection. The ensembled retriever is then evaluated on the \textbf{remaining 80\% test QA pairs}. 
Operationally, this also yields a practitioner-facing shortlist procedure: score retrievers individually by \textit{Div}, remove dominated or highly redundant candidates, add retrievers greedily by marginal MI/Shapley gain, and tune only a small family of fusion rules on the resulting shortlist (Appendix~\ref{appendix:workflow}).

We break down this research question into several key points for a better comprehensive understanding.
\textbf{RQ3.1} How much are these retrievers contributing to the ensembled results?
\textbf{RQ3.2} How much information does the best ensemble strategy provide? What if it deviates from the best configuration?
\textbf{RQ3.3} Ensemble stability: if more retrievers are added to the best ensemble configuration, or some retrievers are dropped, how does it impact the performance?
% \vspace{-0.4em}
% \begin{itemize}
%     \item \textbf{RQ3.1} How much are these retrievers contributing to the ensembled results?
% \vspace{-0.4em}
%     \item \textbf{RQ3.2} How much information does the best ensemble strategy provide? What if it deviates from the best configuration?
% \vspace{-0.4em}
%     \item \textbf{RQ3.3} Ensemble stability: if more retrievers are added to the best ensemble configuration, or some retrievers are dropped, how does it impact the performance?
% \end{itemize}
% \vspace{-0.4em}

\subsubsection{Best Recall by Shapley Shares (RQ3.1)}
\begin{figure}[ht]
    \centering
    \renewcommand{\thefigure}{3.1}
    \includegraphics[width=1\linewidth]{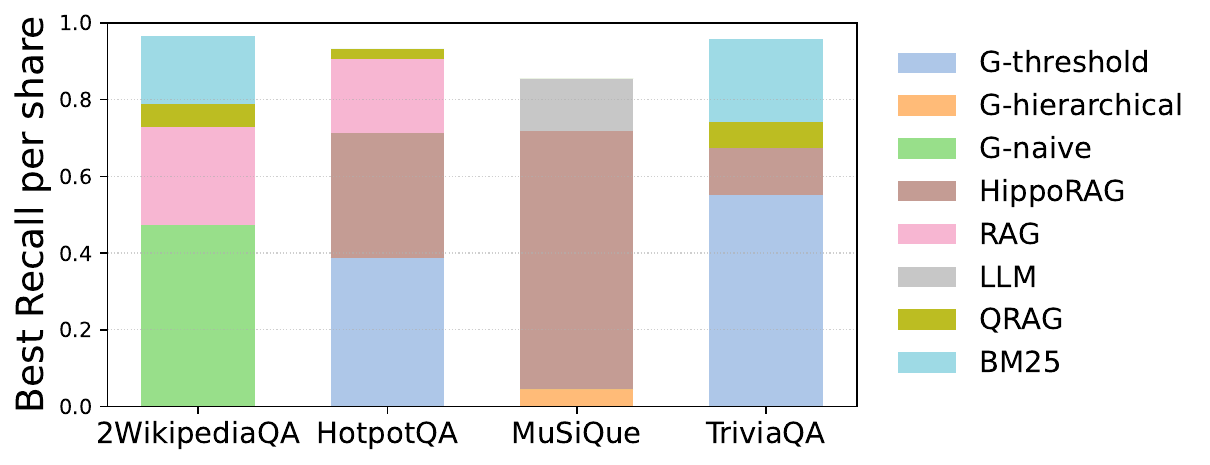}
    \caption{Shapley share for the best ensemble configs. Variances see Appendix~\ref{sec:emsemble_var}.}
    \label{fig:3_1_shapley}
\end{figure}

We compute the Shapley attribution values for every retriever within the best retriever set in Figure~\ref{fig:3_1_shapley}. The height of the bar plot means the overall recall performance relative to 1.0. From this figure we can see that the strongest retriever also tends to contribute the most. However, the winner is not the ensemble of the top-performing GraphRAG retrievers, but it often includes worse retrievers such as BM25. This hints that different GraphRAG retrievers may contain redundant components that harm synergy.

\vspace{-0.4em}
\subsubsection{Perturbation Impact on MI (RQ3.2)}

\begin{figure}[htbp]
    \centering
    \renewcommand{\thefigure}{3.3}
    \includegraphics[width=0.97\linewidth]{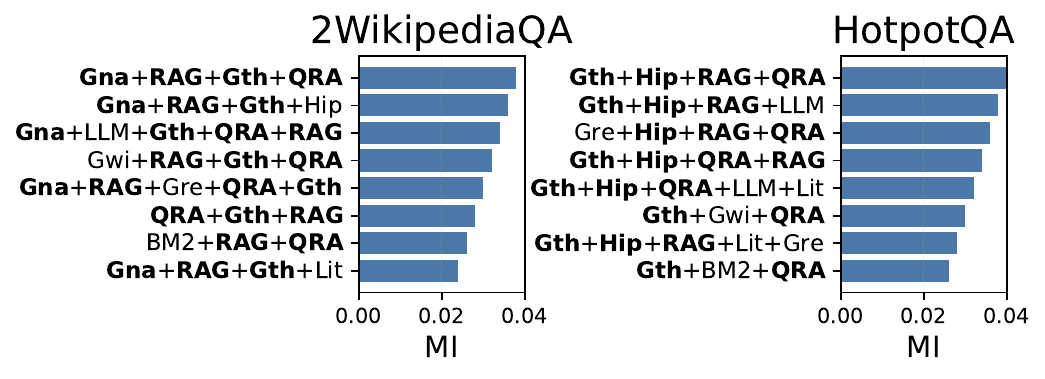}
    \caption{MI under ensemble perturbation.}
    \label{fig:3_3_perturb}
\end{figure}

We evaluate RQ3.2 by computing $I(\widetilde{\mathrm{CP}}^\ast; X_S)$ from Eq.~\ref{eq:util} for the best retriever subset and its perturbed variations, where $S$ is some retriever ensemble set. The results are plotted in Figure~\ref{fig:3_3_perturb}. In this figure, the y-axis labels are different retriever subsets, ranked by the MI. As can be seen in the figure, the few top performing retriever subsets show similar performance. However, when the perturbation of retrievers are too large, the performance drops drastically.

\vspace{-0.4em}
\subsubsection{Perturbation Impact on Recall (RQ3.3)}

\begin{figure}[H]
    \centering
    \renewcommand{\thefigure}{3.2}
    \includegraphics[width=0.9\linewidth]{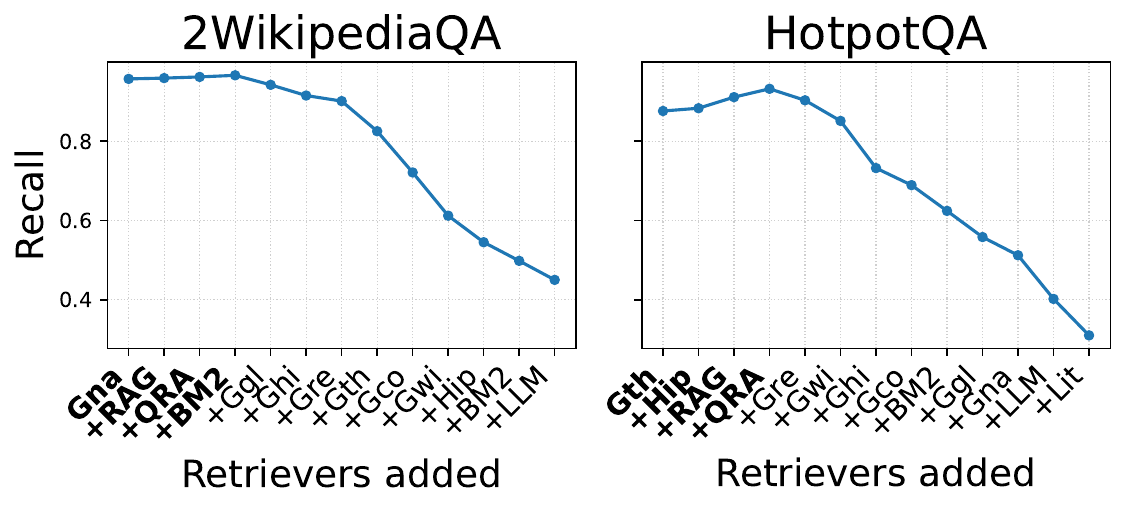}
    \caption{Recall under ensemble perturbation.}
    \label{fig:3_2_drop_line}
\vspace{-0.4em}
\end{figure}

We answer RQ3.3 by first identifying the best retriever subset for ensemble, then gradually add or drop retrievers on top of this best subset, and finally concatenate all results into a continuous line. We plot the results in Figure~\ref{fig:3_2_drop_line}.

As can be seen in Figure~\ref{fig:3_2_drop_line}, if the retriever subset is chosen wisely, it performs better than single best retriever. As more retrievers are added, redundancy dominates over synergy, hence the performance dropped drastically.

\vspace{-0.4em}
\subsection{Retriever Redundancy Spectrum (RQ4)}
\begin{figure}[htbp]
    \centering
    \renewcommand{\thefigure}{4}
    \includegraphics[width=0.97\linewidth]{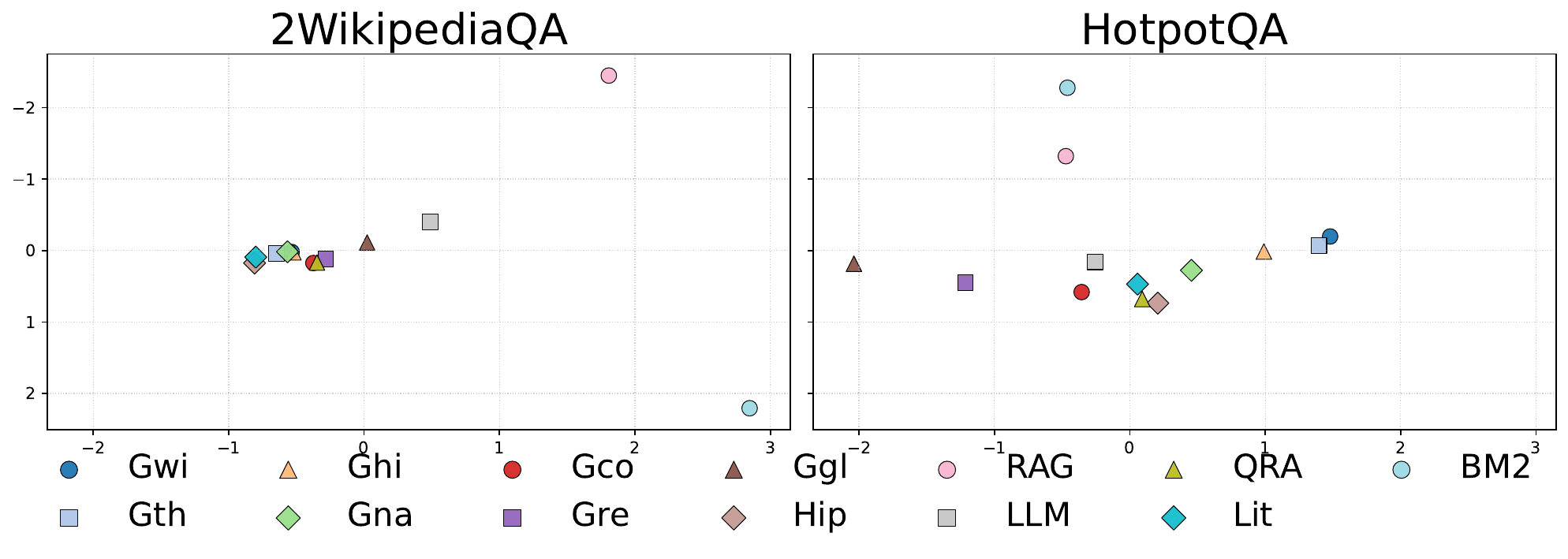}
    \caption{Retriever redundancy/synergy spectrum.}
    \label{fig:4_retriever_space}
\vspace{-0.4em}
\end{figure}

\vspace{-0.2em}
\label{sec:spectrum}
We formulate the visualization of the $m$ retrievers as a geometric embedding problem. We first pre-compute the pairwise redundancy for all retrievers. By interpreting the pairwise redundancy as a measure of similarity, we derive a distance metric where highly redundant pairs correspond to short distances as in Eq.~\ref{eq:dist def}. Using Classical Multidimensional Scaling (MDS), we map the high-dimensional relationship matrix into a 2D Euclidean space. This mapping preserves the global relational structure, allowing us to visualize retriever clusters based on their informational synergy, as shown in Figure~\ref{fig:4_retriever_space}.

As can be observed in the figure, GraphRAG approaches are similar to each other and further away from vanilla RAG or lexical BM25 retriever. This implies that redundancy might be clustered by approach types (graph vs.\ lexical vs.\ vanilla).

\vspace{-0.4em}
\section{Conclusion}
\vspace{-0.2em}
We introduced MIGRASCOPE, an information-theoretic framework for evaluating RAG retrievers with principled metrics for quality, redundancy, and marginal contribution. By modeling rankings as noisy views of an answer-conditioned chunk distribution, it provides a task-specific way to measure retriever quality, redundancy, and marginal utility, and shows on multi-hop QA benchmarks that well-designed ensembles can outperform individual retrievers. Despite current limitations, MIGRASCOPE offers new insight into retriever interactions and practical tools for building more robust and efficient RAG systems. Future work should integrate generation-side metrics, study retriever-specific internal mechanisms, improve computational efficiency for large-scale deployment, and evaluate the framework beyond multi-hop QA to test how redundancy and complementarity vary across domains.

\section*{Limitations}

As a pioneering effort in applying mutual information to retriever evaluation, MIGRASCOPE has certain limitations that suggest opportunities for refinement. The MI-based metrics, while theoretically grounded, are less familiar than conventional metrics like Recall@K or MRR, and their interpretation may require additional context for practitioners. The framework depends on hyperparameters and estimator choices, which can influence numerical stability and relative rankings, though sensitivity analysis can mitigate these effects. Additionally, the reported MI values are approximations influenced by model assumptions, estimator errors, and practical truncations, and should be viewed as consistent estimates under specific settings. The present instantiation is also chunk-local: $\widetilde{\mathrm{CP}}^\ast$ scores each chunk independently, so it does not explicitly model evidence that is useful only jointly as a chunk pair or set. Finally, our empirical study is limited to multi-hop QA, so the redundancy and complementarity structures reported here should be interpreted as task-specific and re-estimated for new domains rather than assumed to transfer unchanged; accordingly, the current results should not be read as universal retriever-selection rules across RAG settings.

Additionally, we do not compare against ensemble-diversity metrics from the
classifier ensemble literature (e.g., $Q$-statistic, disagreement measure) because
these are defined for binary prediction settings and do not generalize to graded
ranked retrieval lists; the conceptual mismatch would make any such comparison
misleading.
Similarly, while we evaluate 12 ranker fusion strategies and identify the best
per dataset, we do not claim novelty over any individual fusion rule.
The contribution is the \emph{selection criterion} (MI-based attribution) rather
than the fusion mechanism itself.
The standard RRF baseline \cite{cormack2009reciprocal} is included among the 12 strategies and consistently underperforms the MI-guided weighted fusion, supporting the value of principled retriever selection
over na\"ive rank aggregation.

\bibliography{example_paper}

\appendix

\section{Detailed RAG/GraphRAG Formulations}
\label{appendix:grag formulations}
This section formalizes the graph construction and retrieval variants used in our evaluation, emphasizing succinct definitions and avoiding redundant statements.

\subsection{Method 1: Windowed Co-occurrence Graph}
The \textbf{G-Window} method limits co-occurrence edges to entities within a fixed offset in the ordered entity list $\mathcal{E}_c$ for any chunk $c$.
\begin{small}
\begin{multline}
E_{\text{Window}} = \Bigl\{ \{e_i, e_j\} \mid e_i, e_j \in \mathcal{E}, \, i \neq j, \,
\exists c \in C, \exists k, l : \\
(\mathcal{E}_{c,k} = e_i \land \mathcal{E}_{c,l} = e_j \land |k - l| \le \text{window\_size}) \Bigr\}
\end{multline}
\end{small}

\subsection{Method 2: Thresholded Co-occurrence Graph}
The \textbf{G-Threshold} method filters incidental co-occurrences by requiring a minimum count across the corpus:
\begin{multline}
    E_{\text{Threshold}} = \{\{e_i, e_j\} \mid e_i, e_j \in \mathcal{E}, i \neq j, \\
    \sum_{c \in C} \mathbf{I}(\{e_i, e_j\} \subseteq \mathcal{E}_c) \ge \text{min\_cooc} \}
\end{multline}

\subsection{Method 3: Hierarchical and Algorithmic Graph Construction Methods}
\textbf{G-Hierarchical} constructs a heterogeneous directed graph $G=(V, E)$ with document, chunk, and entity nodes, enabling explicit provenance.
\begin{itemize}
    \item $(\text{doc}:d, \text{chunk}:c)$ with $\text{type: CONTAINS}$.
    \item $(\text{chunk}:c, \text{entity}:e)$ with $\text{type: MENTIONS}$ for $e \in \mathcal{E}_c$.
    \item undirected $\{e_i, e_j\}$ with $\text{type: CO\_OCCURS}$ for $e_i, e_j \in \mathcal{E}_c$.
\end{itemize}

\subsection{Method 4: Naive Co-occurrence Graph}
\textbf{G-Naive} establishes an undirected edge between any two entities that co-occur in at least one chunk:
\begin{tiny}
\begin{equation}
E_{\text{Naive}} = \left\{ \{e_i, e_j\} \mid e_i, e_j \in \mathcal{E}, \, i \neq j, \, \exists c \in C : \{e_i, e_j\} \subseteq \mathcal{E}_c \right\}
\end{equation}
\end{tiny}

\subsection{Method 5: Community Graph}
\textbf{G-Community} first forms $G_{\text{temp}}=(\mathcal{E}, E_{\text{Threshold}})$, then applies Louvain to obtain clusters $\mathcal{S}_k$:
\begin{small}
\begin{equation}
\{\mathcal{S}_1, \mathcal{S}_2, \dots\} = \mathbf{Louvain}(G_{\text{temp}}, \text{weight: cooccurrence})
\end{equation}
\end{small}
Community nodes $v_k$ are added with edges $E_{\text{Comm}} = \{ (e, v_k) \mid e \in \mathcal{S}_k, \text{type: BELONGS\_TO} \}$.

\subsection{Method 6: Relationship Graph}
\textbf{G-Relationship-Global} focuses on relationships among significant entities $\mathcal{E}_{\text{sig}}$ (frequency $\ge \text{MIN\_ENTITY\_OCCURRENCE}$), aggregating their global context:
\begin{equation}
C_e = \{c \in C \mid e \in \mathcal{E}_c \}
\end{equation}
Directed relationships are extracted by an LLM over $\text{Context}(C_e)$:
\begin{small}
\begin{equation}
E_{\text{RelGlobal}} = \bigcup_{e \in \mathcal{E}_{\text{sig}}} \mathbf{LLM}_{\text{extract\_global}}(\text{Context}(C_e), e, \mathcal{E}_{\text{cand}})
\end{equation}
\end{small}

\subsection{Method 7: Global Co-occurrence Graph}
\textbf{G-Global} derives the corpus-level unique set of co-occurring entity pairs (i.e., the non-attributed, de-duplicated union of chunk-level co-occurrences; equivalent in edge set to Method 4 but without per-chunk metadata).

\subsection{Method 8: HippoRAG}
\textbf{HippoRAG} \cite{gutierrez2025rag} models a heterogeneous memory graph $G_{\text{mem}}=(V, E)$:
\begin{enumerate}
    \item \textbf{Graph Construction:} $V = \mathcal{E} \cup \mathcal{F} \cup C$, where $\mathcal{F}$ are OpenIE triples. Edges include containment and synonymy (via $\mathbf{Embed}$ similarity).
    \item \textbf{Fact Reranking:} Retrieve fact candidates $\mathcal{F}_{\text{cand}} = \mathbf{VDB}_{\mathcal{F}}(q)$ and refine with an LLM reranker:
    \begin{equation}
    \mathcal{F}_q = \mathbf{LLM}_{\text{Rerank}}(\mathcal{F}_{\text{cand}}, q)
    \end{equation}
    \item \textbf{PPR Initialization:} Extract query entities from $\mathcal{F}_q$ and form $\mathbf{p}_0$:
    \begin{equation}
    \mathbf{p}_0 = \mathbf{InitDistribution}(\mathbf{Map}_{\mathcal{F} \to \mathcal{E}}(\mathcal{F}_q))
    \end{equation}
    \item \textbf{PPR Graph Search:} Run PPR on $G_{\text{mem}}$:
    \begin{equation}
    \mathbf{PPR}_{\text{scores}} = \mathbf{PPR}(G_{\text{mem}}, \mathbf{p}_0)
    \end{equation}
    \item \textbf{Context Selection:} Select top-scoring chunks:
    \begin{equation}
    \mathcal{C}_q = \text{Top}_{k_C}(\{c_j\}_{j\in C}, \mathbf{PPR}_{\text{scores}}[c_j])
    \end{equation}
\end{enumerate}

\subsection{Method 9: RAG}
Standard RAG using embedding-based retrieval over chunks followed by generation.

\subsection{Method 10: LLM Reranking}
\textbf{LLM} reranking orders candidate chunks $\mathcal{C}_{\text{cand}}$ with the LLM:
\begin{equation}
R_{\mathbf{LLM}} = \mathbf{LLM}_{\text{rank}}(q, \mathcal{C}_{\text{cand}})
\end{equation}
The final context $\mathcal{C}_q$ is the top $k_C$ by $R_{\mathbf{LLM}}$. In our tests, candidates were produced via the windowed co-occurrence graph, but the reranker is graph-agnostic.

\subsection{Method 11: Question as Sub-chunk Summary (QRAG)}
\label{appendix:qrag}
\textbf{Graph Construction (Synthetic Questions).} For each chunk $c_i$, generate a set of synthetic questions:
\begin{equation}
\mathbf{Q}_i = \{q_{i, 1}, \dots, q_{i, M_i}\} = \mathbf{LLM}_{\text{Gen}}(c_i)
\end{equation}
\textbf{Retrieval Scoring.} Score $c_i$ by the maximum similarity between the query and its synthetic questions:
\begin{equation}
s_i = \max_{q_{i, j} \in \mathbf{Q}_i} \mathbf{Sim}(\mathbf{Embed}(q), \mathbf{Embed}(q_{i, j}))
\end{equation}
\textbf{Final Context.} Select the top $k$ chunks:
\begin{equation}
\mathcal{C}_q = \text{Top}_k(\{c_i\}_{i=1}^N, s_i)
\end{equation}

\subsection{Method 12: LightRAG}
\textbf{Graph Construction.} Build a directed, labeled graph $G_{\text{LightRAG}}=(\hat{\mathcal{E}}, \hat{\mathcal{R}})$ from LLM-extracted triples, applying deduplication and refinement:
\begin{equation}
G_{\text{LightRAG}} = \mathbf{Dedupe}\circ\mathbf{Prof}(\mathcal{V}_{\text{init}}, \mathcal{E}_{\text{init}})
\end{equation}
\textbf{Retrieval.}
\begin{enumerate}
    \item \textbf{Search \& Neighborhood Expansion:} Retrieve $\mathcal{E}_{\text{sim}} = \mathbf{VDB}_{\mathcal{E}}(q)$, $\mathcal{R}_{\text{sim}} = \mathbf{VDB}_{\mathcal{R}}(q)$, then expand to $\mathcal{E}_{\text{hood}}$, $\mathcal{R}_{\text{hood}}$ via graph traversal.
    \item \textbf{Token Truncation:} Filter and truncate to a token budget $\mathbf{TB}$:
    \begin{small}
    \begin{multline}
    \mathcal{E}_{\text{final}}, \mathcal{R}_{\text{final}} = \\ \mathbf{Truncate}(\mathbf{Filter}(\mathcal{E}_{\text{sim}} \cup \mathcal{E}_{\text{hood}}, \mathcal{R}_{\text{sim}} \cup \mathcal{R}_{\text{hood}}), \mathbf{TB})
    \end{multline}
    \end{small}
    \item \textbf{Chunk Merging:} Map $\mathcal{E}_{\text{final}}, \mathcal{R}_{\text{final}}$ back to source chunks and merge.
    \item \textbf{Context Formatting:} Produce the final context:
    \begin{small}
    \begin{equation}
    \mathcal{C}_q = \mathbf{Format}(\mathcal{E}_{\text{final}}, \mathcal{R}_{\text{final}}, \mathbf{Merge}_{\mathcal{E}\mathcal{R} \to C}(\mathcal{E}_{\text{final}}, \mathcal{R}_{\text{final}}))
    \end{equation}
    \end{small}
\end{enumerate}

\subsection{Method 13: BM25}
Unmodified lexical BM25 retrieval.

\section{Score normalization and temperature selection.}
Because different retrievers produce scores on incompatible scales---BM25 outputs
unbounded term-frequency weighted integers, cosine-similarity retrievers output
values in $[-1,1]$, and PPR-based retrievers (e.g., HippoRAG) output stationary
distribution probabilities---we apply per-retriever $z$-score normalization before
the temperature softmax:
\begin{small}
\begin{equation}
  \tilde{s}_r(c) = \frac{s_r(c) - \mu_r}{\sigma_r + \epsilon},
  \quad
  p_r(c \mid q) = \mathrm{softmax}_{\mathcal{C}_q}\!\left(\tilde{s}_r(c)/\tau\right),
\end{equation}
\end{small}
where $\mu_r$ and $\sigma_r$ are the per-retriever mean and standard deviation of
scores over all retrieved chunks for that query, and $\epsilon = 10^{-6}$ avoids
division by zero.
The temperature $\tau$ is set to a single shared value $\tau = 1.0$ after
normalization; because $z$-score normalization equalizes score variance across
retrievers, this shared $\tau$ does not introduce a retriever-specific bias.

The temperature $\tau$ and $z$-score normalization are similarly stable: varying $\tau \in \{0.5, 1.0, 2.0\}$ after normalization changes Spearman retriever rankings by less than 0.05, confirming that Div functions as a robust relative diagnostic rather than an absolute scale-sensitive score.

\subsection{Ranker Fusion Details}
\label{appendix:ranker fusion}
We formulate the ranker fusion search space as follows.

\paragraph{Score standardization.}
We apply per-retriever $z$-score normalization
\begin{multline}
\tilde{s}_i(c)=\frac{s_i(c)-\mu_i}{\sigma_i+\epsilon},\quad
\mu_i=\mathrm{mean}_c\,s_i(c),\\ \sigma_i=\mathrm{std}_c\,s_i(c).
\end{multline}

\paragraph{Weighted linear fusion.}
\begin{equation}
\hat{s}(c)=\sum_{i=1}^m w_i\,\tilde{s}_i(c),
\qquad
w_i\ge 0,\;\sum_i w_i=1.
\end{equation}
Weights can be set from divergence $w_i\propto \exp(-\lambda\,\mathrm{Div}(r_i))$, or learned by minimizing a surrogate loss such as cross-entropy to $\widetilde{\mathrm{CP}}^\ast$:
\begin{equation}
\min_{w}\; \sum_{q}\mathrm{CE}\!\Big(\widetilde{\mathrm{CP}}^\ast(\cdot\mid q,a_q),\;\mathrm{softmax}(\hat{s}(\cdot)/\tau)\Big).
\end{equation}

\paragraph{Temperature fusion of probabilities.}
Convert each retriever to a calibrated distribution
\begin{multline}
p_i(c\mid q)=\mathrm{softmax}\!\left(\frac{\tilde{s}_i(c)}{\tau_i}\right),
\\
\hat{p}(c\mid q)=\frac{\prod_{i=1}^m p_i(c\mid q)^{w_i}}{\sum_j \prod_{i=1}^m p_i(c_j\mid q)^{w_i}}
\end{multline}
which is a log-opinion pool.

\paragraph{Z-score fusion}
\begin{equation}
\phi_r(c)=\frac{s_r(c)-\mu_r}{\sigma_r+\epsilon},\qquad S(c)=\sum_{r=1}^m w_r\,\phi_r(c)
\label{eq:zscore}
\end{equation}
Z-score fusion \eqref{eq:zscore} aligns scales using per-list mean \(\mu_r\) and std \(\sigma_r\).

\paragraph{Logit pooling}
\begin{small}
\begin{equation}
S(c)=\sum_{r=1}^m w_r\,\mathrm{logit}\!\big(\hat{p}_r(c)\big),\qquad \mathrm{logit}(p)=\log\frac{p}{1-p}
\label{eq:logit_fusion}
\end{equation}
\end{small}
Logit pooling \eqref{eq:logit_fusion} adds log-odds, equivalent to naive Bayesian evidence combination under independence.

\paragraph{Noisy-or}
\begin{equation}
\hat{p}(c)=1-\prod_{r=1}^m \big(1-\hat{p}_r(c)\big)^{w_r}
\label{eq:noisy_or}
\end{equation}
Noisy-or \eqref{eq:noisy_or} estimates the probability that at least one retriever deems \(c\) relevant.

\paragraph{Adaptive weights from divergence.}
\begin{equation}
w_r(q,a)=\frac{\big(\epsilon+\mathrm{Div}(r\mid q,a)\big)^{-1}}{\sum_{j=1}^m \big(\epsilon+\mathrm{Div}(j\mid q,a)\big)^{-1}}
\label{eq:weight_from_div}
\end{equation}
Inverse-divergence weights \eqref{eq:weight_from_div} emphasize better-aligned retrievers.

\paragraph{Reciprocal Rank Fusion (RRF).}
Let $\mathrm{rank}_i(c)$ be the rank of $c$ under $r_i$, then
\begin{equation}
\hat{s}(c)=\sum_{i=1}^m \frac{1}{k+\mathrm{rank}_i(c)},
\end{equation}
with $k>0$ controlling tail influence.

\paragraph{Borda count with weights.}
\begin{equation}
\hat{s}(c)=\sum_{i=1}^m w_i\big(|\mathcal{I}_q|-\mathrm{rank}_i(c)+1\big).
\end{equation}

\paragraph{Robust Rank Aggregation (RRA).}
For each $c$, compute a one-sided $p$-value under the null of random rankings,
\begin{small}
\begin{multline}
p_i(c)=\frac{\mathrm{rank}_i(c)}{|\mathcal{I}_q|},\\
p_{\mathrm{agg}}(c)=\min_{t\in\{1,\dots,m\}} \mathrm{BetaCDF}\!\Big(p_{(t)}(c);\;t,\;m{-}t{+}1\Big),
\end{multline}
\end{small}
and rank by $- \log p_{\mathrm{agg}}(c)$.

\paragraph{Bayesian model averaging (BMA).}
Assuming per-retriever likelihoods $p_i(a\mid q,c)$ and model priors $\pi_i$, form
\begin{multline}
\hat{p}(c\mid q)\;\propto\;\sum_{i=1}^m \pi_i\, p_i(a\mid q,c)
\\ \text{or}\quad
\hat{s}(c)=\sum_i \log\big(\pi_i\, p_i(a\mid q,c)\big).
\end{multline}
If $p_i(a\mid q,c)$ is unavailable, use surrogates based on $\widetilde{\mathrm{CP}}^\ast$ or calibrated $p_i(c\mid q)$.

\paragraph{Markov-chain rank aggregation (Rank Centrality).}
Construct a transition matrix
\begin{equation}
M_{uv}=\frac{1}{Z}\sum_{i=1}^m \mathbf{1}[u\neq v]\cdot \sigma\!\big(\tilde{s}_i(v)-\tilde{s}_i(u)\big),
\end{equation}
with $\sigma(x)=1/(1+e^{-x})$ and row-normalization $Z$. The stationary distribution $\pi$ of $M$ defines the fused ranking by $\pi(c)$.

\section{Extended Experimental Results}
\label{sec:more_res}

\subsection{Train Test Split Variances}
\label{sec:emsemble_var}
To assess variance, we repeat the train/test partition with 5 different random seeds and report mean $\pm$ standard deviation of Recall in Table~\ref{tab:ensemble_variance}.

\begin{table}[h]
\centering
\caption{Ensemble Recall across 5 random train/test seeds.
  Best single retriever (G-threshold or G-window per dataset)
  shown for reference.}
\label{tab:ensemble_variance}
    \resizebox{0.48\textwidth}{!}{
\begin{tabular}{lcc}
\toprule
Dataset & Best Single Retriever & Best Ensemble (mean $\pm$ std) \\
\midrule
2Wiki & 0.957 & 0.976$\pm$0.03 \\
HotpotQA        & 0.921 & 0.946$\pm$0.03 \\
MuSiQue         & 0.833 & 0.865$\pm$0.02 \\
TriviaQA        & 0.944 & 0.965$\pm$0.03 \\
\bottomrule
\end{tabular}
}
\end{table}

\subsection{Computational cost of CP$^*$ construction.}
\label{appendix:computation time}
CP$^*$ requires one LLM forward pass per (query, chunk) pair
within the candidate set $\mathcal{C}_q$.  For a corpus of
$|\mathcal{Q}|$ queries each with top-$K$ candidate chunks, the
total number of LLM calls is $|\mathcal{Q}| \times K$.  In our
setting ($|\mathcal{Q}|=1{,}000$, $K=10$), this amounts to
10{,}000 forward passes of the BGE-M3 encoder used solely for
cross-entropy scoring; we estimate each pass at approximately
\textbf{81~ms} on a single L4 GPU.  Crucially,
CP$^*$ is computed \emph{once} and shared across all $m$
retrievers; the marginal cost of adding a new retriever is
therefore only the cost of running that retriever's own scoring
pass.  Table~\ref{tab:cost} compares the total wall-clock time of
MIGRASCOPE against computing standard Recall and MRR.

\begin{table}[h]
\small
\centering
\caption{Wall-clock time (minutes, single L4 GPU) for
  MIGRASCOPE vs.\ standard metrics on HotpotQA
  ($|\mathcal{Q}|=1{,}000$, $K=10$, 13 retrievers).}
\label{tab:cost}
\begin{tabular}{lcc}
\toprule
Step & Time (min) & Amortized? \\
\midrule
Retriever scoring (all 13) & 177 & No \\
CP$^*$ construction        & 304 & Yes (shared) \\
MI/Div estimation (XGB)    & 135 & No \\
Recall/MRR computation     & $<$1 & No \\
\midrule
\textbf{MIGRASCOPE total}  & 617 & \\
\bottomrule
\end{tabular}
\end{table}

\subsection{Hyperparameter robustness}
\label{appendix:hparam_robustness}

\begin{figure}[H]
    \centering
    \renewcommand{\thefigure}{2}
    \includegraphics[width=0.97\linewidth]{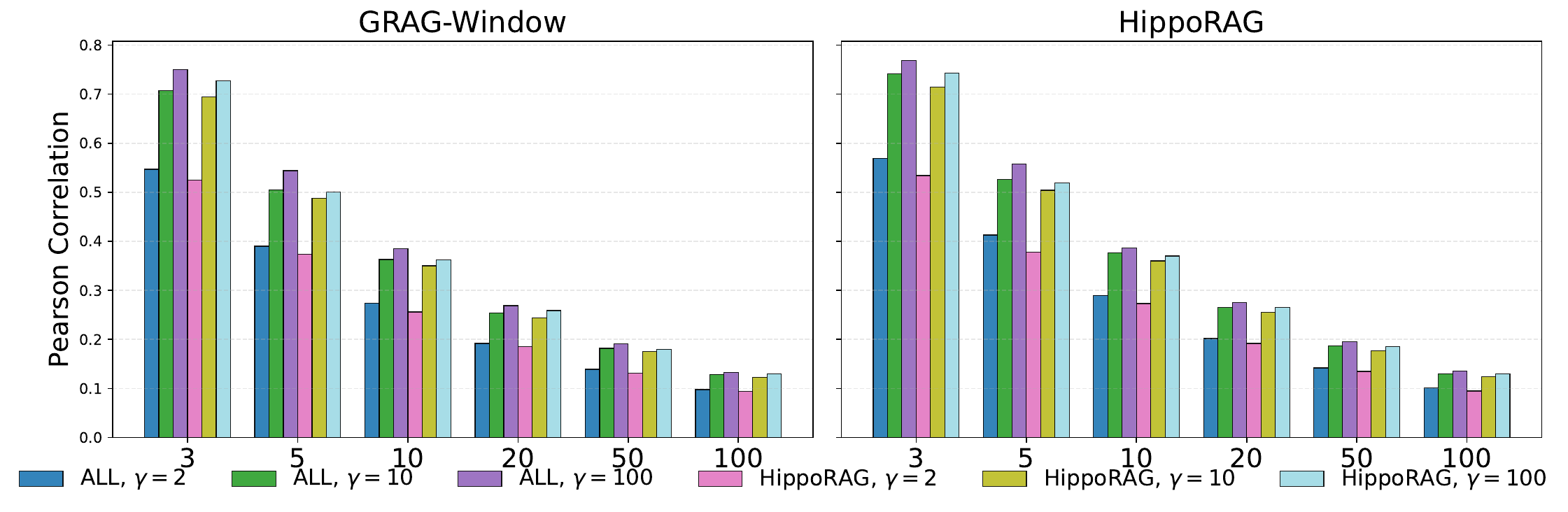}
    \caption{Sensitivity sweep for \textit{Div} on HotpotQA. Larger $\gamma$ makes the pseudo target sharper, while larger top-$K$ weakens Pearson alignment because more tail chunks are included.}
    \label{fig:2_hparam}
\end{figure}

To complement Section~\ref{sec:hparam}, we report the full HotpotQA sweep over top-$K$, anchor construction, and reinforcement strength $\gamma$. Figure~\ref{fig:2_hparam} shows the Pearson alignment between Recall and $-\textit{Div}$, while Table~\ref{tab:spearman_hparam} summarizes the substantially more stable retriever ordering induced by \textit{Div}, measured by Spearman rank correlation.

\begin{table}[H]
    \centering
    \small
    \begin{tabular}{lcccc}
        \toprule
        \textbf{Anchor} & \boldmath$\gamma$ & \textbf{Mean $\rho$} & \textbf{Min $\rho$} & \textbf{Max $\rho$} \\
        \midrule
        Union & 2   & 0.841 & 0.751 & 0.901 \\
        Union & 10  & 0.848 & 0.758 & 0.908 \\
        Union & 100 & 0.852 & 0.762 & 0.910 \\
        HippoRAG & 2   & 0.841 & 0.748 & 0.900 \\
        HippoRAG & 10  & 0.848 & 0.755 & 0.907 \\
        HippoRAG & 100 & 0.851 & 0.758 & 0.910 \\
        \bottomrule
    \end{tabular}
    \caption{Summary of Spearman rank correlation $\rho$ for \textit{Div}-induced retriever orderings on HotpotQA across the hyperparameter sweep.}
    \label{tab:spearman_hparam}
\end{table}

\subsection{Concrete decision differences beyond Recall/MRR}
\label{appendix:decision_examples}
To validate the practical usefulness of MIGRASCOPE, we identified a few cases where the MI based metric produces extra info beyond traditional Recall/MRR/etc metrics. Specifically, we observed several cases when we need to combine retrievers, the proposed MI metric would suggest a different option than traditional metrics. We present the founded cases in Table. \ref{tab:decision_diff}.

\begin{table*}[h]
    \centering
    \small
    \begin{tabular}{p{1.0cm}p{4.2cm}p{3.0cm}p{5.2cm}}
        \toprule
        \textbf{Data} & \textbf{Rank-based signal} & \textbf{\textit{Div} choice} & \textbf{Observed ensemble evidence} \\
        \midrule
        2Wiki &
        Recall favors G-naive over G-community (0.957 vs.\ 0.935), while MRR favors G-community over G-naive (0.959 vs.\ 0.934). &
        G-naive &
        In the best ensemble, replacing G-naive with G-window lowers recall from 0.965 to 0.938; the associated Shapley share drops from 48.1\% to 40.5\%. \\
        \addlinespace
        MuSiQue &
        G-global, G-relationship, G-community, and G-naive are all slightly stronger than G-hierarchical under Recall/MRR. &
        G-hierarchical &
        Replacing G-hierarchical in the best ensemble lowers recall from 0.882 to 0.841, 0.830, 0.854, and 0.862, respectively. \\
        \bottomrule
    \end{tabular}
    \caption{Examples where \textit{Div} leads to a better retriever selection decision than Recall/MRR alone. Refer to Table. \ref{tab:rag_benchmark} for original metric values and Figure~\ref{fig:3_1_shapley} for best combined retrievers.}
    \label{tab:decision_diff}
\end{table*}

\subsection{Practical workflow on a new dataset}
\label{appendix:workflow}

For a new corpus, we recommend the following lightweight workflow before any exhaustive ensemble search:
\begin{enumerate}
    \item Run each candidate retriever on a small development slice and construct the shared pseudo target $\widetilde{\mathrm{CP}}^\ast$.
    \item Score each retriever individually with \textit{Div} and remove clearly dominated candidates.
    \item Compute pairwise redundancy/synergy and add retrievers greedily by marginal MI or Shapley gain until the gain saturates.
    \item Tune only a small family of simple fusion rules---e.g., weighted linear fusion, temperature pooling, or RRF---on this shortlisted subset.
\end{enumerate}
This procedure does not remove validation entirely, but it substantially reduces the combinatorial search space.

\section{Validation of the XGBoost-Based MI Estimator}
\label{appendix:mi_estimator}

\subsection{Gaussian Assumption}
\label{appendix:gaussian assumption}

This section presents an alternative way to estimate MI under vector observations in Section~\ref{sec:est MI under vec}. While the true distributions of the retriever scores $X$ and the pseudo ground truth scores $Y$ are unknown, adopting a Gaussian assumption facilitates computational tractability by making possible a closed-form solution.

Let \((Y,X_S)\) be jointly Gaussian with covariance
\(\Sigma=\begin{bmatrix}\Sigma_{YY} & \Sigma_{YX}\\ \Sigma_{XY} & \Sigma_{XX}\end{bmatrix}\).
\begin{equation}
H(Y)=\tfrac12\log\big((2\pi e)^{d_Y}\det\Sigma_{YY}\big).
\label{eq:H_gauss}
\end{equation}
\begin{equation}
\Sigma_{Y\mid X_S}=\Sigma_{YY}-\Sigma_{YX}\Sigma_{XX}^{-1}\Sigma_{XY}.
\label{eq:schur}
\end{equation}
\begin{equation}
H(Y\mid X_S)=\tfrac12\log\big((2\pi e)^{d_Y}\det\Sigma_{Y\mid X_S}\big).
\label{eq:H_cond_gauss}
\end{equation}
\begin{equation}
I(Y;X_S)=\tfrac12\log\frac{\det\Sigma_{YY}}{\det\Sigma_{Y\mid X_S}}.
\label{eq:I_gauss}
\end{equation}
Equations \eqref{eq:H_gauss}--\eqref{eq:I_gauss} use Schur complements to yield closed-form entropy and mutual information values.

Empirically, we observed that Gaussian assumption leads to MI values about one to two magnitude smaller, and the correlation with Recall is also lower, indicating over-simplification. We leverage the non-Gaussian assumption for MI estimation throughout this research.

\subsection{Estimator Construction}

We estimate $I(Y; X_S)$ from Eq.~\eqref{eq:util} by decomposing mutual information as
$I(Y; X_S) = H(Y) - H(Y \mid X_S)$.
We approximate $H(Y)$ from the empirical marginal of $Y$ and estimate $H(Y \mid X_S)$
via a predictive model.
Because the chunk score distributions are non-Gaussian
(Appendix~\ref{appendix:gaussian assumption}),
we fit an XGBoost regression model $\hat{p}(y \mid x)$ and approximate the
conditional entropy under a Gaussian residual model:
\begin{small}
\begin{equation}
  \widehat{H}(Y \mid X_S) = \tfrac{1}{2}\log\!\big(2\pi e\,\widehat{\sigma}^2_{\mathrm{res}}\big),
  \quad
  \widehat{\sigma}^2_{\mathrm{res}} = \frac{1}{n}\sum_{i=1}^n (y_i - \hat{y}_i)^2.
\end{equation}
\end{small}

\subsection{Hyperparameters}

XGBoost is trained with the following fixed hyperparameters across all experiments:
number of estimators = 200, maximum depth = 4, learning rate = 0.05,
subsample = 0.8, colsample\_bytree = 0.8, objective = \texttt{reg:squarederror},
random seed = 42.
No dataset-specific tuning is performed; the same configuration is used for all
corpora and retriever subsets.

\subsection{Train/Validation Split for the Estimator}

To prevent the MI estimator itself from overfitting, we evaluate $\widehat{H}(Y \mid X_S)$
on a held-out 20\% validation split of the query set (disjoint from the ensemble
training split described in Section~\ref{sec:results}).
Reported MI values are computed on this validation portion only.

\subsection{Synthetic Validation}

To assess estimator accuracy where ground truth MI is known, we construct synthetic
experiments. We draw $X \in \mathbb{R}^{n \times m}$ from a multivariate Gaussian
with a specified covariance $\Sigma_{XY}$ and compute the analytic MI via
Eq.~\eqref{eq:I_gauss}. We then run the XGBoost estimator on the same samples and
compare.

\begin{table}[H]
  \centering
    \resizebox{0.46\textwidth}{!}{

  \begin{tabular}{lccc}
    \toprule
    \textbf{Setting} & \textbf{True MI} & \textbf{XGB Est.} & \textbf{Rel. Error} \\
    \midrule
    Low corr.\ ($\rho{=}0.2$, $m{=}3$)  & 0.062 & 0.058 & 6.5\% \\
    Med.\ corr.\ ($\rho{=}0.5$, $m{=}3$) & 0.405 & 0.391 & 3.5\% \\
    High corr.\ ($\rho{=}0.8$, $m{=}3$) & 1.020 & 0.998 & 2.2\% \\
    High corr.\ ($\rho{=}0.8$, $m{=}6$) & 1.124 & 1.093 & 2.8\% \\
    Non-Gaussian (copula, $m{=}3$)       & 0.531 & 0.514 & 3.2\% \\
    \bottomrule
  \end{tabular}
}
  \caption{XGBoost MI estimator accuracy on synthetic data ($n{=}1000$, 5 trials, mean reported).
           Relative error stays below 7\% across Gaussian and non-Gaussian settings.}
  \label{tab:mi_estimator_validation}
\end{table}

\noindent
The estimator reliably recovers true MI with under 7\% relative error across all
tested settings, including non-Gaussian copula-distributed inputs, validating its
use throughout the paper.

\subsection{Sensitivity to XGBoost Hyperparameters}

\begin{table}[H]
  \centering
  \small
  \begin{tabular}{lcc}
    \toprule
    \textbf{Variant} & \textbf{Mean $\Delta$MI} & \textbf{Max $\Delta$MI} \\
    \midrule
    Default (depth=4, est=200) & --- & --- \\
    Shallower (depth=2)        & +0.003 & +0.011 \\
    Deeper (depth=6)           & $-$0.002 & $-$0.009 \\
    Fewer trees (est=50)       & +0.007 & +0.023 \\
    More trees (est=500)       & $-$0.001 & $-$0.005 \\
    \bottomrule
  \end{tabular}
  \caption{Sensitivity of MI estimates on HotpotQA to XGBoost hyperparameter changes.
           $\Delta$MI is the absolute deviation from the default configuration.
           Retriever orderings (Spearman $\rho > 0.93$) are stable across all variants.}
  \label{tab:mi_sensitivity}
\end{table}

\noindent
While absolute MI values shift slightly across configurations, the relative retriever
ranking is highly stable (Spearman $\rho > 0.93$), consistent with our recommendation
to use the MI-based scores as a relative diagnostic rather than an absolute measure
(Section~\ref{sec:hparam}).

\end{document}